\documentclass[aps,twocolumn,prb]{revtex4}

\usepackage[dvipdfmx]{graphicx}

\newcommand{\nnm}{\nonumber}

\newcommand{\ra}{\rangle}
\newcommand{\la}{\langle}

\newcommand{\upa}{\uparrow}
\newcommand{\dna}{\downarrow}

\begin{document}

\title{Readout using Resonant Tunneling in Silicon Spin Qubits}%

\author{Tetsufumi Tanamoto}
\affiliation{Department of Information and Electronic Engineering, Teikyo University,
Toyosatodai, Utsunomiya 320-8511, Japan} 
\email{tanamoto@ics.teikyo-u.ac.jp}

\author{Keiji Ono}
\affiliation{Advanced device laboratory, RIKEN, Wako-shi, Saitama 351-0198, Japan} 

\begin{abstract}
Spin qubit systems are one of the promising candidates for quantum computing.
The quantum dot (QD) arrays are intensively investigated by many researchers.
Because the energy-difference between the up-spin and down-spin states is very small, 
the detection of the qubit state is of prime importance in this field.
Moreover, many wires are required to control qubit systems.
Therefore, the integration of qubits and wires is also an important issue.
In this study, the measurement process of QD arrays is theoretically investigated 
using resonant tunneling, controlled by a conventional transistor.
It is shown that the number of possible measurements during coherence time can exceed a hundred under 
the backaction of the measurements owing to the nonlinear characteristics of resonant tunneling. 
It is also discussed to read out the measurement results by the conventional transistor.
\end{abstract}

\maketitle

\section{Introduction}
Silicon quantum computers are among the most popular topics in the fields of quantum physics and engineering.
The legacy of integrated circuit technologies is expected to be inherited by silicon qubit systems.
Currently, one-dimensional qubit arrays have been successfully 
fabricated~\cite{Hensgens,Mills,Fedele,Veldhorst1,Ha,Vandersypen}.
Silicon qubits use narrowly confined carriers with their interactions governed by short-range spin--spin interactions.
As the distance between two qubits increases, exchange coupling decreases exponentially. 
Accordingly, the control of distant qubits becomes much more difficult 
than that of the nearest neighboring qubits. 
Moreover, the number of wires required to manage a qubit is approximately greater than five~\cite{Yoneda}.
Therefore, a sufficiently small qubit is required. 
Because the wiring structure at the bottom layer should have the same dimension as the qubit, 
the tight structure is inevitable~\cite{Veldhorst}.
In order to connect the distant qubits, additional mechanism is necessary 
such as the shuttling qubits~\cite{Noiri1,Utsugi},
or the mediated electrons~\cite{Marcus,Croot}. 
Shuttling qubit systems require overhead to precisely control electron shuttling.
Therefore, an extra area is required compared with mediated electrons.

In Ref.\cite{TanaAIP}, we have proposed a common gate structure using mediated electrons 
which are directly embedded between conventional transistor channels. 
However, since most investigations thus far have been conducted in QD systems ~\cite{Hensgens,Mills,Fedele,Veldhorst1,Ha},
a mediated electron system in QD array should be considered presently.
Not all QDs in QD systems have to be qubits; some of them can act as measurement devices.
In Ref.~\cite{TanaPRB}, we have theoretically investigated the effect of the measurement
of QD system by using the Green's function method~\cite{Jauho}, 
and we proved that the energy-difference of two QDs, which corresponds to the 
qubit energy-difference under magnetic fields, becomes detectable for 
appropriate coupling between the QDs and the electrodes. 
However, in Ref.~\cite{TanaPRB}, we only considered a small-bias region with a fixed-value resistor, 
while Zeeman splitting in channel QDs is ignored.
In this study, we investigate a side-QD system including Zeeman splitting of all QDs. 
Because there are many literature regarding the quantum operations 
in several QDs~\cite{Fei,Medford,Burkard,Kandel,Noiri},  
our emphasis is on the measurement properties.

Generally, QD with electrodes exhibits resonant tunneling in a wide bias region. 
The specific features of resonant tunneling include a nonlinear current 
with coherent tunneling~\cite{Capasso,Fujita,Medford,Kim2}, 
and the transport properties  have been discussed in literature~\cite{Reed,Jiang}.
In this study, we theoretically show that, by focusing on the nonlinear characteristics 
of resonant current, efficient measurements against the backaction of the measurements can be obtained.
Specifically, we propose a system to measure the energy difference between two qubits using the nonlinear change in current depending on the spin direction.
We illustrate that the number of the detections of the spin states during decoherence 
exceeds a hundred, which can lead to the possibility of surface coding in the measurement phase~\cite{Fowler}.

As aforementioned, an advantage of silicon qubits is that they can be directly integrated into conventional complementary metal-oxide-semiconductor (CMOS) circuits on the same chip. 
Therefore, some parts on the chip should connect qubits and CMOS transistors using local wires.
In this study, we investigate the transport properties of the aforementioned qubit system
directly connected to the conventional transistors~\cite{BSIM}. 
A transistor-connected qubit system offers several advantages to control the qubit system. 
In general, the variations in the size of QDs are inevitable,
resulting in the variations of energy levels of the QDs.
The transistor can adjust the difference of the energy levels of the qubits.
Moreover, as the switching speed of a conventional CMOS is in the order of ps~\cite{Jena}, 
the spin qubits can be quickly controlled by the CMOS~\cite{TanaCMOS},
compared with the previous methods of separating the measurement apparatus 
from the cryogenic refrigerator using long wires~\cite{ISSCC1,ISSCC2}.
As semiconductor technologies continue to advance daily and the era of 3D circuits is on the rise~\cite{Chung,Tsutsui,Kim,Ryckaert,Tanaka,Lee,Lue},
active usage of the advances in state-of-the-art technologies is necessary.
Hence, the implementation of the proposed qubit system has been discussed as a near-future structure.

The rest of this study is organized as follows.
In Section~\ref{sec:model}, our proposed model is explained in detail with the formalism using 
the standard Green function method. 
In Section~\ref{sec:results}, the numerical results of our method are presented.
In Section~\ref{sec:discussions}, discussions regarding our results are provided.
Section~\ref{sec:conclusions} summarizes and concludes this study.
In appendix, additional explanations are presented including a 
possible qubit structure using the advanced transistor system.

\begin{figure}
\centering
\includegraphics[width=8.8cm]{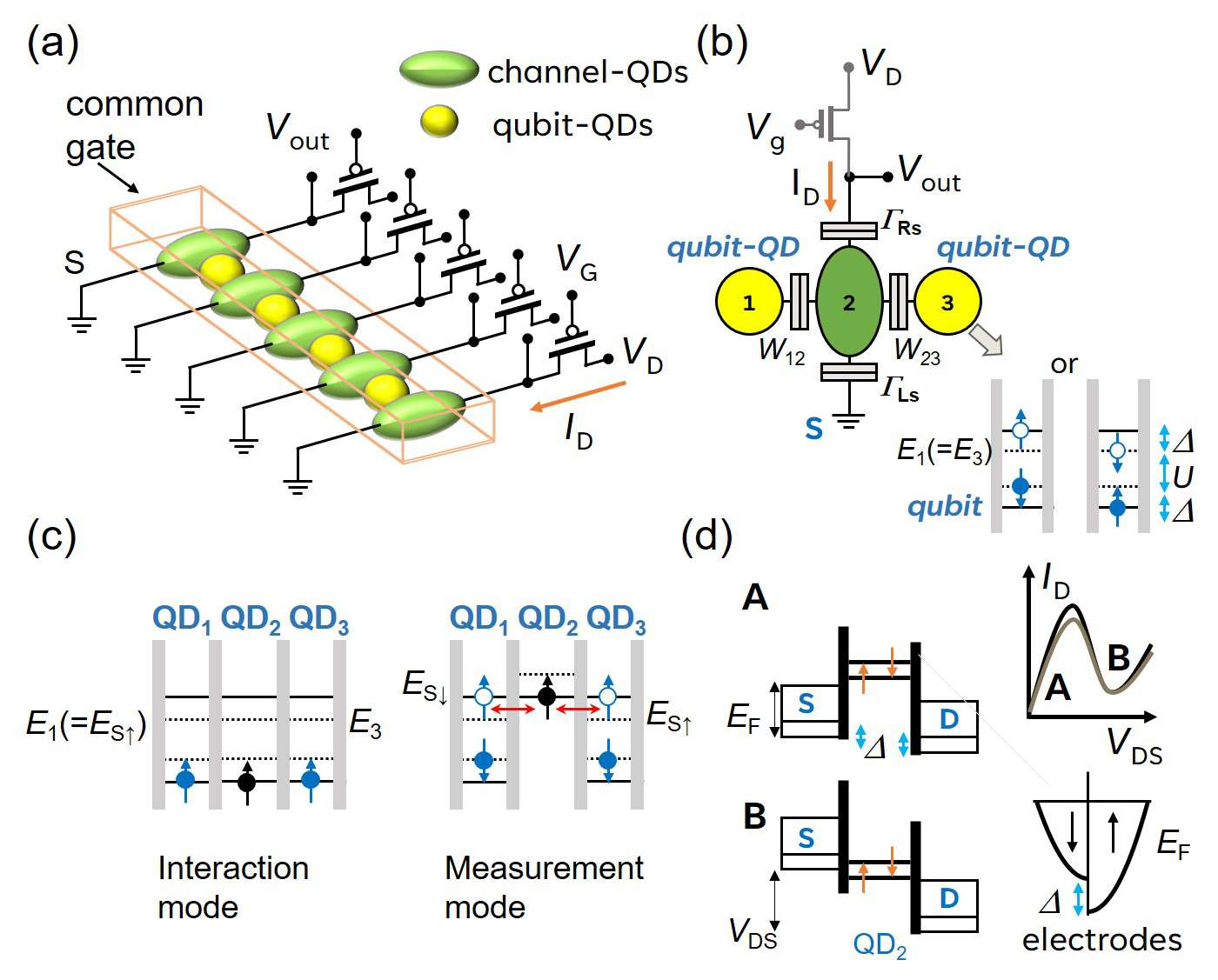}
\caption{Quantum dot (QD) system considered in this study.
(a) Array of coupled qubit system connected to transistors.
The electric potentials of the qubits (yellow circles) are
controlled by a common gate. 
The measurement current flows from the transistors to the channel-QD (green ellipse).
(b) Measurement unit of two qubits using three QDs. 
$\Gamma_{\alpha s}$ denotes the tunneling couplings between the electrodes ($\alpha=L,R$, $s=\pm 1/2$),
and $W_{ij}$s are the coupling strengths between the QDs.
In the measurement mode, the current interacts with 
the upper energy levels of the qubit-QD (right-side figure).
$\Delta$ and $U$ denote Zeeman splitting energy and on-site Coulomb energy, respectively.
$E_i$ denotes an energy-level of the down-spin ($i=1,2,3$).
(c)
Depending on the position of the energy level of the center-QD, 
two modes are available: interaction ($V_D=0$) and measurement ($V_D\neq 0$) modes. 
The black and blue circles denote the electrons of center-QD and qubits, respectively. 
The right figure represents one of the four measurement mode types (Fig.2).  
(d) Typical current-voltage characteristics of resonant tunneling.
Owing to a finite magnetic field, 
the bottom of the Fermi energy of the up-spin current (down-spin current) is shifted downwards (upward).
With an increase in voltage, the current increases (point ``A''), and 
when the bottom of the Fermi level of the drain is above the resonant level, 
the current begins to decrease (point ``B''). 
}
\label{fig1}
\end{figure}

\section{Model}\label{sec:model}
Figure~\ref{fig1}(a) shows the proposed qubit array connected to CMOS transistors.
Qubit-QDs and channel QDs are coupled by tunneling oxides such as SiO${}_2$.
The channel QDs are connected to the transistors.
Figure~\ref{fig1}(b) illustrates the concrete model estimated in this study,
which consists of two QDs designated qubits and a channel QD.
In the qubit-QD, an electron exists at the lowest energy level and works as a spin qubit, as proposed in Ref.~\cite{TanaAIP}.
The next-higher energy level of the qubit-QD is placed 
over the on-site Coulomb interaction ($U$) to form a singlet state ($S$), as shown in Fig.1(c). 
Electrons of the center-QD enter the qubit-QDs only when aligned energy levels of the qubit-QD exist.
If $U$ is estimated as the charging energy, then 
we have $U\propto 46.4$ meV for a QD of 10 nm~\cite{TanaAIP},
which is significantly larger than the operating region $V_D<$ 3meV considered here.
We follow many literature regarding the qubit-qubit operation in three QD system
and focus on the measurement mode here.
Depending on the up and down-spin of the qubits, 
singlet energy levels are represented by $E_1$ and $E_1+\Delta$, respectively. 
($\Delta$ denotes the Zeeman splitting energy).
The energy levels $E_1$ and $E_1+\Delta$ allow only down and up-spin of electrons, respectively, 
as shown in the inset of Fig.1(b).

Measurements are performed when the lowest energy level of the center-QD
is increased to the upper energy levels for the second electron in the qubit QDs, 
such that the current through the center-QD can detect $E_1$ or $E_1+\Delta$, which 
depends on the spin state of the qubits.
The qubit states (spin state of the electron in the lowest energy level of the qubit-QD) 
are indirectly affected by the measurements.

The measurement ($V_D\neq 0$) and interaction ($V_D= 0$) modes are separated by the potential energies of the center-QD, 
as shown in Fig.~1(c).
The energy levels of the qubits are controlled by the common gate, 
while that of the center-QD are controlled by the source-to-drain voltage. 
The energy levels of the qubits are assumed to change independently from that of the center-QD.
In the interaction mode (left side of Fig.~1(c)),
the lowest energy level of the center-QD is reduced such 
that an exchange interaction occurs between the electrons. 
In the measurement mode (right side of Fig.~1(c)), 
the lowest energy level of the center-QD is raised such that 
the qubits do not interact directly with each other through 
electrons in the center-QD.

When a magnetic field is applied, the energy levels of the QDs and electrodes are separated (Fig~\ref{fig1}(d))~\cite{Sanchez}.
The measurement currents of up ($\upa$-current) and down ($\dna$-current) spins 
are assumed to be treated independently by neglecting the spin flips between transports.
Then, the $\upa$-current and $\dna$-current can be treated to have 
different Fermi energies $E_{F\pm}=E_F \pm \Delta/2$.
A negative differential conductance appears when the bottom of the Fermi energy of the source 
exceeds the discrete energy level of the channel QD, as shown in Fig.1(d). 
As both the QDs and electrodes are affected by the Zeeman splitting energy,
the end bias of the negative differential conductance coincides with $\upa$-current and $\dna$-current

\begin{figure}
\centering
\includegraphics[width=8.0cm]{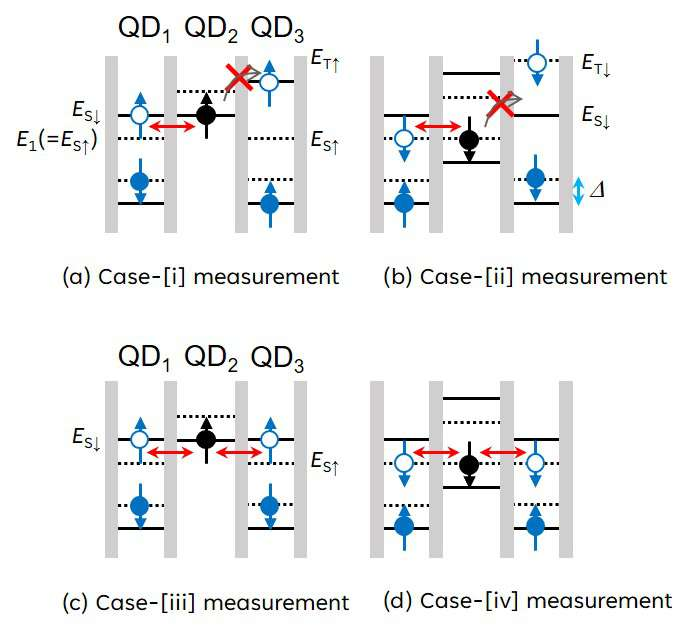}
\caption{
Four current cases in the measurement mode reflect the four qubit states.
The black and blue circles show the electrons of the center-QD and qubits-QDs, respectively. 
The white circle shows the vacant energy level in which other electrons can enter.
$E_S$ and $E_T$ show the singlet and triplet energy-levels, respectively. 
}
\label{figband}
\end{figure}

In the measurement mode (Fig.~\ref{figband}), four cases of energy levels exist in the qubit states on both sides of the center-QD. 
The current of the channel-QDs changes reflecting the different energy levels of the qubit-QDs.
When two qubit-QDs do not exhibit the same spin state (Fig.~\ref{figband}[i] and [ii]), 
the current reflects one of the two qubit states with $E_1$ or $E_1+\Delta$. 
When the two qubit-QDs have the same spin state (Fig.~\ref{figband}[iii] and [vi]), 
the current reflects both the qubit states at the same energy level.
For example, in Fig.~\ref{figband}[i], 
the left and right-qubits have the down and up-spin states, respectively.
The higher energy level of the left-qubit-QD is represented by $E_1+\Delta$, 
while that of the right-qubit-QD is denoted as $E_3(=E_1)$.
Here, the $\upa$-current resonates with the left-qubit, 
but no resonance energy levels of the $\dna$-current between the 
center-QD and the left-qubit exist.
To identify the four cases of the currents ([i]$\sim$[vi]), 
we set the reference current as that of $W_{12}=W_{23}=0$, and 
estimate the current deviations $\Delta I$ 
between the four cases of the currents ([i]$\sim$[vi]) and the reference current.
The reference current is obtained when the energy level of the center-QD exists 
between the lowest and highest energy levels of the qubit-QDs.

As aforementioned, the energy levels of the qubit-QDs are controlled by the common gate 
and that of the center-QDs are controlled by the source and drain voltage. 
Here, the common gate voltage is assumed to be fixed.
Thus, $E_1(=E_3)$ is fixed and $E_2$ changes such that $E_2=E_2^{(0)}+V_D/2$.
Subsequently, cases [i]([iii]) and [ii]([iv]) occur when $E_2=E_1+\Delta$ 
($E_2=E_1-\Delta$).
In addition, $E_1(=E_3) >E_2^{(0)}$ is set such that 
$E_2$ crosses $E_1(=E_3)$ as $V_D$ increases from $V_D=0$.

The Hamiltonian of the QD system depicted in Fig.~\ref{fig1}(b) is given by
\begin{eqnarray}
H&=&\sum_s \sum_{i=1}^3 E_{is} d_{is}^\dagger d_{is}  
\!+\! \!\sum_{\alpha=L,R}\sum_{k_\alpha,s} E_{k_\alpha s} c_{k_\alpha s}^\dagger c_{k_\alpha s}
\nonumber \\ 
&+&\!\sum_{\alpha=L,R}\sum_{k_\alpha,s} 
[V_{k_\alpha,s,2} c_{k_\alpha,s}^\dagger d_{2s}+V_{k_\alpha s}^* d_{2s}^\dagger c_{k_\alpha s} ]
\nonumber \\
&+& \sum_{i=1}^{2}\sum_s W_{i,i+1}(d_{is}^\dagger d_{i+1,s}+h.c.),
\label{Hamilt}
\end{eqnarray}
where $c_{ks}^\dagger$ ($c_{ks}$) creates (annihilates) an electron of momentum $k$ 
and spin $s(=\pm 1/2)$ in the electrodes, 
while $d^\dagger_{is}$ ($d_{is}$) creates (annihilates) 
an electron in the QDs ($i=1,2,3$). 
$E_{k_\alpha s}=E_{k_\alpha}+s g\mu_B B$ is the energy level of the electrode($\alpha=L,R$), 
and $E_{is}=E_{i0}+sg\mu_B B$ is the energy level for three QDs ($i=1,2,3$).

The Zeeman splitting energy is expressed as $\Delta\equiv g \mu_B B$ in the following calculations.
One energy-level in each QD is assumed.
As the Zeeman splitting energy for the channel-QD and electrodes is considered in this study, 
the Hamiltonian is slightly different from that of Ref.~\cite{TanaPRB}.
It is assumed that the current of the up-spin state is independent of that 
of the down-spin state. 
Therefore, the formulation proposed in Ref.~\cite{TanaPRB} can be independently applied to $\upa$-current and $\dna$-current.
The coupling coefficients of the electrodes to the QDs are given by
\begin{equation}
\Gamma_{\alpha,s}(\omega)=2\pi \sum_{k_\alpha} |V_{k_\alpha,s}|^2\delta (\omega-E_{k_\alpha s}),
\end{equation}
($\alpha=L,R$).
Strictly speaking, $\Gamma_{\alpha,s}(\omega)$ depends on the spin direction 
through the density of states. 
However, for the sake of simplicity, we take 
$\Gamma_s\equiv \Gamma_{\alpha,\uparrow}(\omega)=\Gamma_{\alpha,\downarrow}(\omega)(\alpha=L,R)$. 
The current can be expressed by the transmission probability $\mathcal{T}_{2s}(\omega)$, such that
\begin{equation}
I_{D}=\sum_s\frac{e^2}{h}\int d\omega \mathcal{T}_{2s}(\omega) [f_S(\omega)-f_D(\omega) ]
\label{trans},
\end{equation}
where $f_\alpha (\omega)=[\exp[(\omega-\mu_\alpha)/(k_BT)]+1]^{-1}$ ($k_B$, $\mu_\alpha$, and $T$) denote 
the Boltzmann constant, chemical potential of the $\alpha$-electrode, and temperature, respectively (see Appendix~\ref{Gfn})). 
The magnitude of $\Gamma_{\alpha,s}$ is determined such that the 
calculated current is less than a few hundred pA~\cite{Elzerman,Shaji,Lu}. 
In the present setup, the qubit states do not directly interact with the channel electrons.
Thus, it is considered that the effect of the channel current on the coherence of the qubit is weak.
A detailed calculation of the decoherence time $t_{\rm dec}$ is formulated below (Eq.(\ref{tdec})).
As $W_{ij}$ increases, the current reflects the qubit state more strongly.
In the following calculations, $W_{ij}$ is selected such that the number of 
measurements exceeds a hundred.

The core model for $I_D$-$V_D$ of the fin field-effect transistor (FinFET) provided by~\cite{BSIM} 
is considered, given by 
\begin{equation}
I_{\rm D} = \beta (V_{\rm g}-V_{\rm th}-V_{\rm ds}/2)V_{\rm ds}, 
\label{CMOS}
\end{equation}
where $\beta\equiv (\mu {\mathcal W}/L) (\epsilon/EOT)$
($L=1~\mu$m, ${\mathcal W}=80$nm, $\mu=1000$cm$^{2}$V$^{-1}$s$^{-1}$, 
$EOT$=1 nm, and $\epsilon=3.9\times 8.854\times 10^{-12}$ F/m 
represent the gate length, gate width, mobility, oxide thickness, 
and dielectric constant, respectively~\cite{BSIM}).
The transistor characteristics at low temperatures are under an intensive study 
by the CMOS Society~\cite{Pahwa}. 
Therefore, the simple form of Eq.(\ref{CMOS}) is considered.

Numerical methods are applied to solve the equations of the QD device and transistor, 
given by
\begin{equation}
V_D=V_{\rm out}+V_{\rm ds}.
\label{eq1out}
\end{equation}
For a given $V_D$, the current of the channel-QD, expressed by Eq.(\ref{trans}), 
should be equal to that of the transistor, expressed by Eq.(\ref{CMOS}).
Thus, this equation is solved numerically using Newton's method.

The purpose of the measurement is to distinguish between the different spin states 
based on different magnitudes of the measured currents.
Thus, the current difference $\Delta I$ should be larger than the average noise level. 
The possible number of measurements is estimated by 
comparing the decoherence time $t_{\rm dec}$ with the measurement time.
Following Ref.~\cite{Schon}, 
the measurement time $t_{\rm meas}$ is defined using $\Delta I$ and shot noise$S$, 
given by 
\begin{equation}
t_{\rm meas}^{-1}\equiv \frac{(\Delta I)^2}{4S},
\label{measurement}
\end{equation}
Because there are four cases of currents (Fig.2), 
four cases of the measurement times are considered.
The quantum shot noise $S$ is originally expressed by numerous terms, 
and it has a simple form at $T=0$~\cite{Kobayashi,Lopez}.
The classical shot noise of $S=2e\la I \ra$ 
is found to be larger than that in the quantum case.
As shot noise increases, 
the measurement time $t_{\rm meas}$ becomes longer and undesirable. 
Here, the classical form is used as the worst case.

The decoherence time caused by the interactions is described using the golden rule~\cite{Schoelkopf}. 
The last term of the Hamiltonian Eq.~(\ref{Hamilt}) 
is treated as the perturbation term, given by
$H_{\rm int}(t)\equiv \sum_{i=1}^{2}\sum_s W_{i,i+1}(d_{i}^\dagger(t) d_{i+1}(t)+h.c.)$, 
while the non-perturbated terms are those of the current lines ${\rm S}-{\rm QD}_2-{\rm D}$.
Then, $t_{\rm dec}$ can be defined as~\cite{TanaPRB}
\begin{eqnarray}
t_{\rm dec}^{-1} &\approx &\frac{1}{\hbar^2} 
\int_{-\infty}^\infty d\tau e^{-i\omega_{01}\tau} 
\langle H_{\rm int}(\tau)H_{\rm int}(0) \rangle,
\label{tdec} 
\end{eqnarray}
(see Appendix~\ref{Gfn}).
Similar to $t_{\rm meas}$, four cases of $t_{\rm dec}$ corresponding to 
the four currents are considered.

\begin{figure}
\centering
\includegraphics[width=8.5cm]{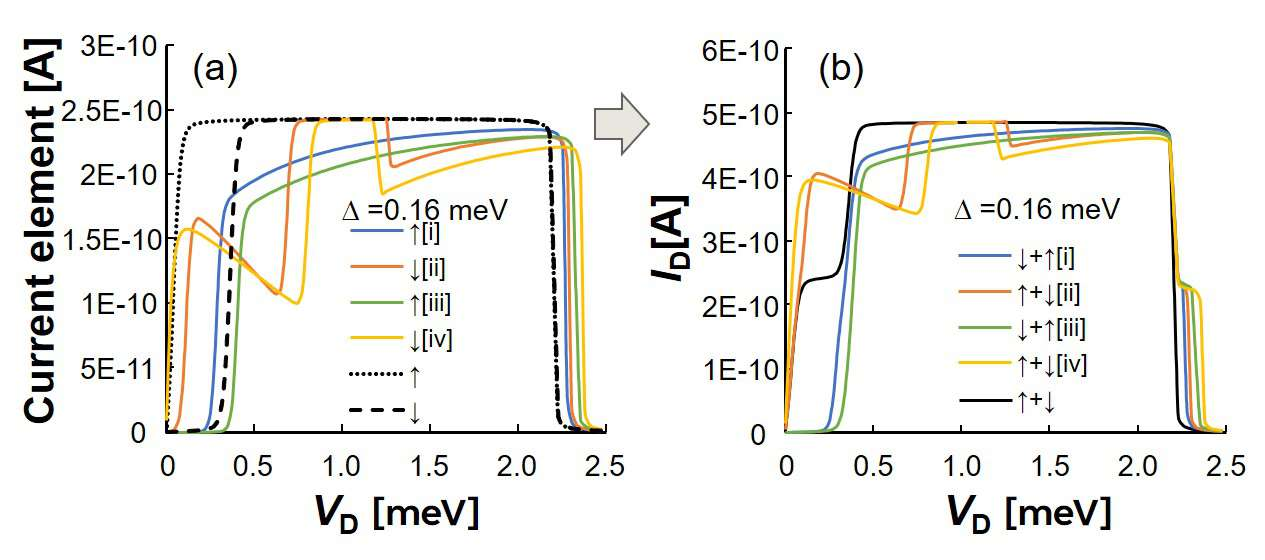}
\caption{
Current characteristics as a function of the applied voltage 
$V_D$ in Fig.1(b) without the transistor.
(a)Individual $\upa$-current and $\dna$-current are shown separately.
(b)Total currents are shown where 
the black solid, dashed, and dotted lines denote the case of $W=0$ (reference current). 
$\Gamma_0=0.02u_0$ and $W=2u_0$ ($u_0=10^{-4}$ eV) is considered for $\Delta=0.16$ meV.
The other solid lines represent the four cases of measurement currents shown in Fig.2.
$E_1=E_3=E_F+2u_0$, $E_2^{(0)}=E_F+u_0$, 
$E_F$=1 meV, and $T=100$ mK.
}
\label{fig3}
\end{figure}

\begin{figure}
\centering
\includegraphics[width=8.5cm]{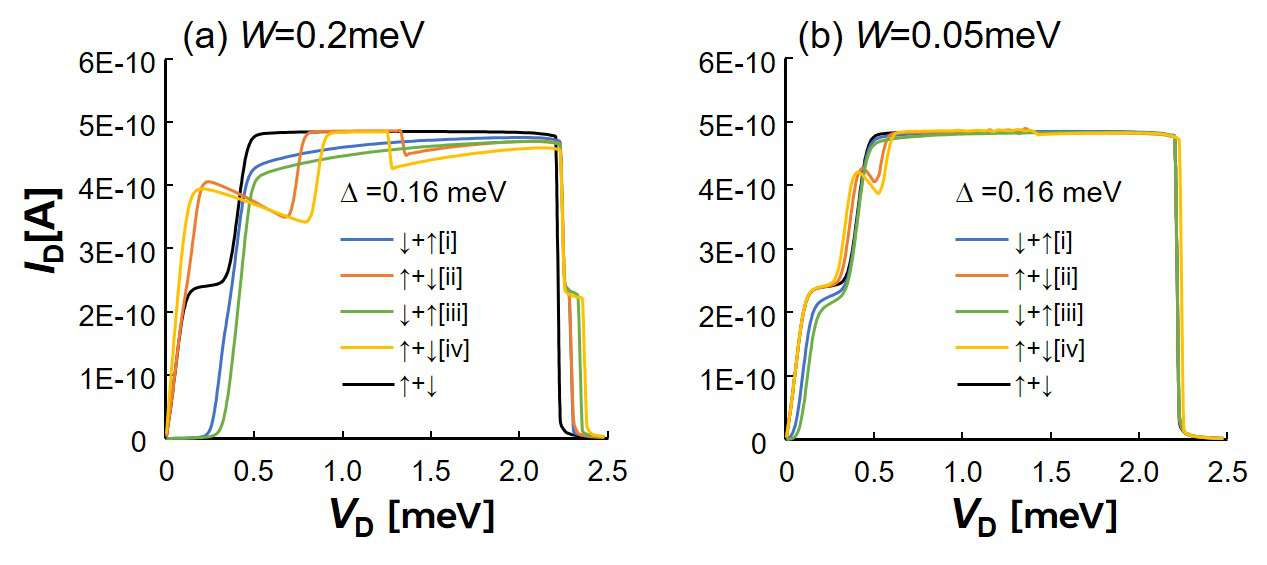}
\caption{
Current characteristics as a function of the applied voltage 
$V_D$ in Fig.1(b) with the transistor.
(a) $W=2u_0$ and (b) $W=0.5u_0$  for $\Delta=0.16$ meV.
Other parameters are the same as those considered for Fig.~\ref{fig3}.
The black solid lines show the case of $W=0$ (reference current),
while the other solid lines show the four cases of measurement currents shown in Fig.2.
}
\label{fig4}
\end{figure}

\begin{figure}
\centering
\includegraphics[width=8.5cm]{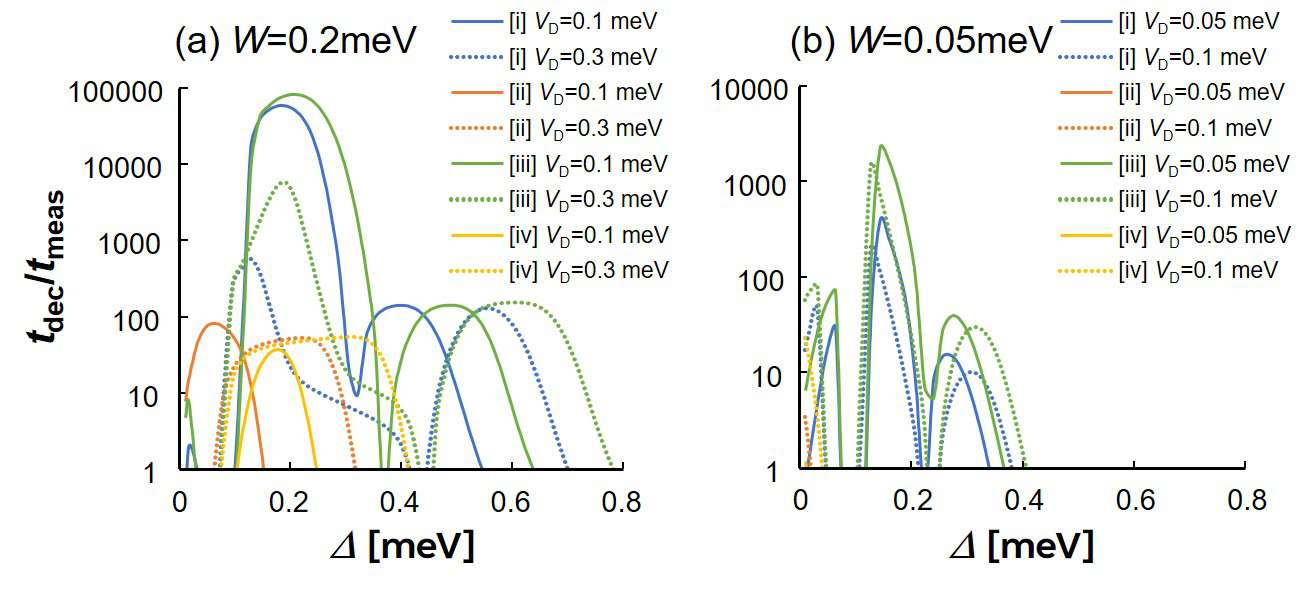}
\caption{
Zeeman splitting energy $\Delta$ dependence of $t_{\rm dec}/t_{\rm meas}$ in $V_D<0.5$meV region.
$t_{\rm dec}/t_{\rm meas}$ is considered to correspond to the approximate number of possible measurements 
before the measurement backaction breaks the coherence. 
(a) $W=2u_0$ and (b) $W=0.5u_0$.
Other parameters are the same as those considered for Fig.~\ref{fig4}.
}
\label{figtmeas}
\end{figure}

\begin{figure}
\centering
\includegraphics[width=8.5cm]{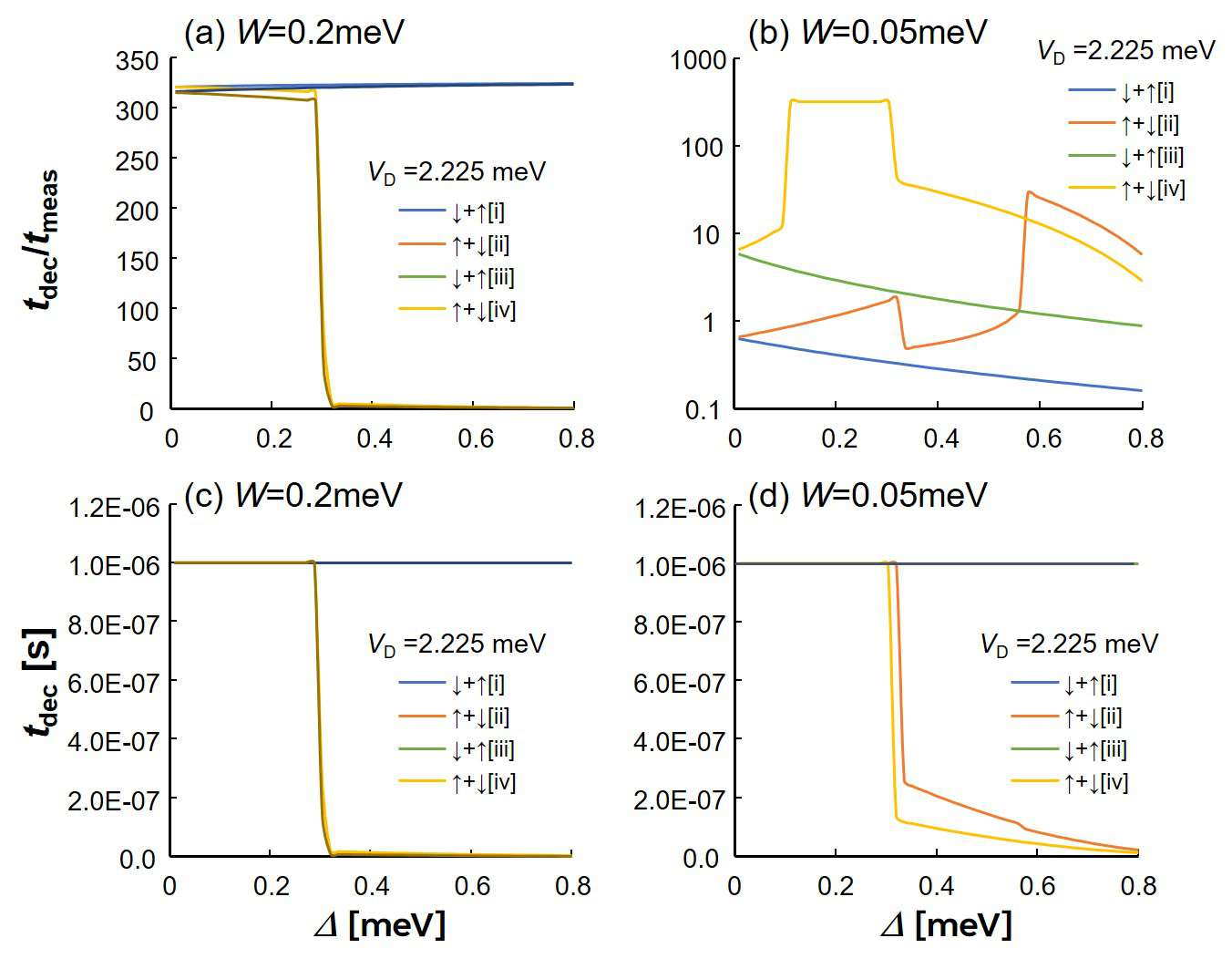}
\caption{
(a)(b)
Zeeman splitting energy $\Delta$ dependence of $t_{\rm dec}/t_{\rm meas}$ at $V_D=2.225$ meV. 
The solid lines present the sum of currents of different spin states. 
(c)(d) Decoherence time $t_{\rm dec}$ calculated using Eq.(\ref{tdec}),
(a)(c) for the cases of [i] and [ii] and (b)(d) for the cases of [iii] and [iv] shown in Fig.2.
The parameters are the same as those of Fig.~\ref{fig4}.
}
\label{figSQD2}
\end{figure}

\section{Numerical Results}\label{sec:results}
Figure~\ref{fig3} shows the $I_D$--$V_D$ characteristics of the system shown 
in Fig.1(b) for different spin states without the transistor. 
The currents increase from $V_D=0$ and decrease from approximately $V_D\sim 2.5$ meV.
Thus, a single resonant peak is observed in the present calculations.
In Fig.~\ref{fig3}(a),  the $\upa$-current and  $\dna$-current
are presented independently, where the dotted line (dashed line) represents the $\upa$-current 
($\dna$-current) for $W_{12}=W_{23}=0$ (reference current).
In Fig.~\ref{fig3}(b), the four cases of channel currents, shown in Fig.2, 
with the reference current have been illustrated.
As an example, the case [i] current consists of the case [i] $\upa$-current in Fig.~\ref{fig3}(a), 
and the $\dna$-spin reference current (dashed line in Fig. ~\ref{fig3}(a)).
The range of $V_D$ is chosen to cover the peak structure of the resonant tunneling.
Cases [i] and [iii], with the $\upa$-currents reflecting the qubit states, show similar behaviors, 
while cases [ii] and [iv], 
with the $\dna$-currents reflecting the qubit states, show similar behaviors.

Energy levels of $E_1=E_3=E_F+2u_0$ and $E_2^{(0)}=E_F+u_0$ 
are selected such that these energy levels are within the range of the 
nonlinear region around $V_D\approx 0$.
The resonant peak ends when $E_2$ reaches $V_D$ such that $E_2(=E_2^{(0)}+V_D/2)=V_D$ 
as shown in Fig.1(d), implying that $V_D \approx 2E_2^{(0)}$.
For $E_F=10 u_0$ and $E_2^{(0)}=E_F+u_0$, 
the peak structure ends at approximately $V_D=22 u_0=2.2$ meV. 
The end of the peaks of the $\upa$-current is the same as that of the $\dna$-current (Fig.1(d)).
On the other hand, the current flow begins  
when $E_2$ is given by $E_2(=E_2^{(0)}+V_D/2)=V_D+E_F$ which leads to $V_D=2(E_2^{(0)}-E_F)$.
For the $\upa$-current ($\dna$-current)
in which the effective Fermi surface is given by $E_F+\Delta/2$ ($E_F-\Delta/2$), 
the bias for the beginning of the current is given by $V_{D\upa}=2(E_2^{(0)}-E_F-\Delta/2)$ 
($V_{D\dna}=2(E_2^{(0)}-E_F+\Delta/2)$).
Thus, the difference of the beginning of the current
is given by $V_{D\upa}-V_{D\dna}=2\Delta E_F=2\Delta$, 
as seen in Fig.~\ref{fig3}.
The current difference $\Delta I$ in Eq.~(\ref{measurement}) is 
calculated as the difference between currents [i]$\sim$[iv] and the reference currents. 
As shown in Fig.~\ref{fig3}, the difference $\Delta I$ 
is prominent around $V_D\sim0$ and $V_D>2.0$ meV.

Figure~\ref{fig4} illustrates the $I_D$-$V_D$ characteristics 
for the system shown in Fig.1(b), with the transistor of gate length $L$=1 $\mu$m and $V_{\rm g}-V_{\rm th}$=0.1V.
By comparing Figs.~\ref{fig4} with Figs.~\ref{fig3}, the $I_D$-$V_D$ characteristics shown in Fig.~\ref{fig4} 
are similar to that shown in Fig.~\ref{fig3}. 
This is because the resistance of the transistor of $L$=1 $\mu$m is considerably smaller than that of the QD system. 
The effect of coupling $W$ for $W=0.05$ meV is smaller than that for the $W=0.2$ meV case.
For a smaller $V_D$ region (left of the resonant peak), 
different effects are observed on the $\upa$-current (Fig.2[i] and [iii]) and $\dna$-current(Fig.2[ii] and [iv]).
Conversely, for a larger $V_D$ region (right of the resonant peak), 
the change from the reference current looks small for $W=0.05$ meV.

A large $\Delta I$ is desirable for a short measurement time $t_{\rm meas}$.
As can be seen in Fig.~\ref{fig4}, the differences between the four currents 
[i]$\sim$[iv] are prominent at $V_D<0.5$meV and 2.2$<V_D<$ 2.5 meV.
We consider that $t_{\rm dec}/t_{\rm meas}$ corresponds to the number of possible measurements.
Figure~\ref{figtmeas} shows $t_{\rm dec}/t_{\rm meas}$ in $V_D<0.5$meV region.
$t_{\rm dec}/t_{\rm meas}$ is found to exceed 100
for the cases [i] and [iii].
However, those of cases [ii] and [iv] are less than 100.
Thus, the region $V_D<0.5$meV is preferable only for the cases [i] and [iii].
Figures~\ref{figSQD2} (a) and (b) show $t_{\rm dec}/t_{\rm meas}$ at $V_D=2.225$ meV. 
More than 100 measurements are expected for all cases [i]$\sim$[iv].
Figures~\ref{figSQD2} (c) and (d) show the corresponding $t_{\rm dec}$ values as a function of $\Delta$.
$t_{\rm dec}$ increases for smaller $\Delta$, 
where the maximum value of $t_{\rm dec}$ is taken as 1$\mu$s, reflecting the 
recent improvement in the decoherence time of spin qubits~\cite{Veldhorst,Kobayashi2}.
Thus, when the limit of the decoherence time is increased, a greater number of measurements is expected. 
Conversely, when the limit of the coherence 
time is decreased, a smaller number of measurements is expected. 
Such a limit of the decoherence time is considered to arise from 
the other origin of the qubit system, which must be determined experimentally. 
The recessed patterns shown in Fig.~\ref{figSQD2}(b) are obtained from 
the abrupt structure of $t_{\rm dec}$, as provided in Ref.~\cite{TanaPRB}.

$t_{\rm dec}/t_{\rm meas}$ varies depending on $E_1$ and $E_2$. 
Guidelines for achieving the optimal $E_2^{(0)}$ and $E_1$ have not been established
other than  that $E_2^{(0)}>E_1$ and $E_2^{(0)}$ is close to $E_1$ 
from the results of the repeated simulations.
$t_{\rm dec}/t_{\rm meas}>100$ for $\Delta=0.25$ meV is feasible for 
$B_z=2.16$ T because 1 meV$\sim$8.64 T. 
If we regard the possibility of more than hundred measurements 
as the less than 1\% measurement error,
the side-QD structure is considered to lead to the future surface code.

\begin{figure}
\centering
\includegraphics[width=8.0cm]{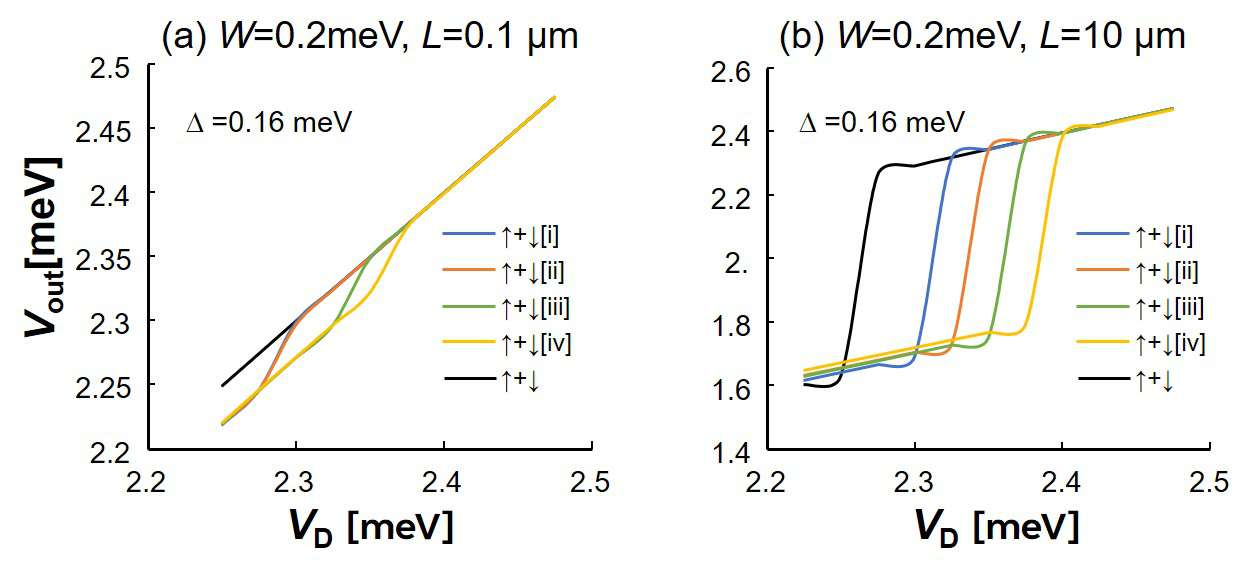}
\caption{
Output $V_{\rm out}$ of the system shown in Fig.1(b), when the 
gate length of the transistor $L$ changes.
$V_{\rm g}-V_{\rm th}=0.1V$.
(a) $L$=1~$\mu$m,  and (b) $L$=10~$\mu$m.
Other parameters are the same as those considered for Fig.~\ref{figSQD2}.
}
\label{figVout}
\end{figure}

Figure~\ref{figVout}(a) shows $V_{\rm out}$ when the gate length of the transistor is $L=1~\mu$m. 
The change of $V_{\rm out}$ is less than 50 $\mu$V 
because the resistance of the channel-QD part ($\sim$ 1 meV /1nA$\sim 10^{6}\Omega$) 
is much larger than that of the transistor of $L=1~\mu$m. 
Figure~\ref{figVout}(b) shows the case of $L=10~\mu$m, where the change in 
$V_{\rm out}$ exceeds 500 $\mu$V. 
This result shows that long channel transistor is required to 
obtain sufficiently large $V_{\rm out}$.
Depending on $V_{\rm g}-V_{\rm th}$, $I_D$-$V_D$ characteristics changes.
Generally, the variations in the size of QDs are inevitable, 
and appropriate bias $V_D$ changes for different channel-QDs.
The direct connection of the transistor to the qubit-system has the 
advantage that variations of $I_D$-$V_D$ characteristics 
can be adjusted by controlling the connected transistor.
A larger $t_{\rm dec}/t_{\rm meas}$ is observed for a larger $V_{\rm g}-V_{\rm th}$,
but with a smaller change in $V_{\rm out}$.
Thus, a trade-off exists between the changes in $V_{\rm out}$ and $V_{\rm g}-V_{\rm th}$. 
Detailed relationships are a topic for future research.
The $I_D$–$V_D$ characteristics and $t_{\rm dec}/t_{meas}$ for $L=10~\mu$m are provided in Appendix ~\ref{Temp}.

\section{Discussion}\label{sec:discussions}
In the present measurement, spin directions are not determined by a single shot. 
Thus, the ratio $t_{\rm meas}/t_{\rm dec}$ does not exactly represent 
the required number of measurements.
The qubit state can be determined by taking an average of repeated measurements.
In any case, a large number of possible measurements will lead to the future surface code.
In previous experiments~\cite{ISSCC1,ISSCC2}, several amplification circuits have been used 
to detect small differences between the up and down-spin states.
Direct connection of the transistor to the qubit is observed to increase the resolution 
of the current differences (see Appendix~\ref{2D}).

The effect of unexpected trap sites in the transistor should be considered~\cite{TanaAIP}. 
The local electric field produced by static traps could be adjusted by changing the transistor voltages. 
The effect of dynamic traps such as random telegraph noise (RTN) 
would be significant  because the voltage fluctuation is in the order of mV.
Owing to the long interval of RTN ($\sim \mu$s),  
repeated measurements might be desirable to identify the effect of RTN.

Although Eq.~(\ref{tdec}) shows the first-order effect 
of tunneling between the qubit-QD and center-QD, 
the decoherence time is a complicated 
function of the energy levels of QDs as discussed in Ref.~\cite{TanaPRB}.
The ratio $W/\Gamma_0$ is larger than that in Ref. ~\cite{TanaPRB}, 
indicating that the electrons have to spend some time in the QD region 
to obtain information about the qubits.
When higher-order tunneling, such as cotunneling effects, is included, 
the formula presented in Eq.~(\ref{tdec}) becomes significantly more complicated, 
and is a topic of future research.

Note that, in the cases of [i] and [ii], which of QD1 and QD3 has larger energy is not determined. 
What is known by the current detection for the cases of [i] and [ii]
is that the two qubits have different spin states.
The absolute energy of each qubit-QD is determined by detecting the edge qubits
because the corresponding current reflects only one qubit.
The edge qubit is described such that either $W_{12}$ or $W_{23}$ is zero.
In this study, we focused on the case for $W_{12}\neq 0$ and $W_{23}\neq 0$, 
because the edge qubits behave similarly.

In this study, we did not mention quantum operations.
Regarding the quantum operations between the qubits, 
several previous proposals~\cite{Fei,Medford,Burkard,Kandel,Noiri} would be applied.
In these quantum operations, a magnetic-field gradient is required, 
and it is better that the micromagnets are placed on the structure of Fig.1(a).

As an extension of Fig.1(a), the stacked structure can be further considered 
based on a nanosheet which is one the near-future three-dimensional (3D) semiconductor systems~\cite{Chung,Tsutsui,Kim,Ryckaert,Tanaka,Lee,Lue}(see Appendix ~\ref{3D}).
However, the fabrication of nanosheets itself requires advanced fabrication technologies. 
This is also a future issue.

In the present setup, the qubit state is indirectly affected by the channel current. 
To protect the qubit state more efficiently, 
addition of ancilla qubit between the channel-QD and qubit-QD can be considered (Appendix ~\ref{FiveQD}). 
A detailed analysis will be conducted in the future.

\section{Conclusions}\label{sec:conclusions}
To summarize, the measurement process for several QD arrays is theoretically investigated using resonant tunneling. 
Moreover, direct connections of the qubit system to the conventional transistor are considered for reading measurements.
By using the nonlinear region of resonant tunneling phenomenon, 
the small energy difference between the qubits is enhanced, 
resulting in the possibility of more than hundreds of measurements during the decoherence time. 
By connecting a conventional transistor to the qubits, the measurement results are directly transmitted 
to the conventional CMOS circuits without long wires. 
Although the proposed structure is expected to be an extension of state-of-the-art semiconductor technology, 
a concrete fabrication process will be investigated in the future. 
Since the proposed system has several parameters, 
the determination of optimal parameters will depend on future experiments.

\subsection*{DATA AVAILABILITY}
The data that supports the findings of this study are available within the article.

\begin{acknowledgments}
We are grateful to S. Takagi, T. Mori and H. Fuketa for fruitful discussions.
This work was partly supported by MEXT Quantum Leap Flagship Program (MEXT Q-LEAP), 
Grant Number JPMXS0118069228, Japan.
This work was also supported by JSPS KAKENHI Grant Number JP22K03497.
\end{acknowledgments}

\subsection*{AUTHOR CONTRIBUTIONS}
T.T. conceived and designed the theoretical calculations. 
K.O. discussed the results from viewpoint of experimentalist. 

\subsection*{CONFLICT OF INTEREST}
The authors have no conflicts to disclose.

\appendix

\section{Green's Function Method}\label{Gfn}
\subsection{$I_D$-$V_D$ Characteristics}
Green's functions are derived using the equation-of-motion method~\cite{Jauho}.
Here, the formulation has been briefly explained.
The time-dependent behavior of the operator $d_{i}(t)$ is derived from $i\hbar \frac{d~d_{i}(t)}{dt}=[H,d_{i}(t)]$, 
and we have
\begin{eqnarray}
\omega d_{i}(\omega) = [H, d_{i}(\omega)].
\label{eqmotion}
\end{eqnarray}
Generally, the current through the QD system is described by the transport properties of the left and right electrodes. 
Hereafter, the notations $L(=S)$ and $R(=D)$ have been used. 
The current $I_{Ls}$ for spins $s$ between the central and left electrodes 
is given by
\begin{eqnarray}
I_{Ls}(t)
&=& \frac{e}{\hbar} \int dE {\rm Re} \left\{
\sum_{k_L} V_{k_Ls,i}G^<_{d_{is},c_{ik_L s}}(E)
\right\}, \label{currentformula}
\end{eqnarray}
where 
$G^<_{d_{is},c_{ik_\alpha s}}(t,t')\equiv i \langle c_{ik_\alpha s}^\dagger(t') d_{is}(t)\rangle$, 
and
$G^<_{c_{ik_\alpha s},d_{is}}(t,t)=-[G^<_{d_{is},c_{ik_\alpha s}}(t,t)]^*$. 
The total current is assumed to be conserved between the source (left electrode) and drain (right electrode).
Thus, the left electrode currents $I_{L}$ and the right electrode currents $I_{R}$ 
satisfy the relation $I_L=-I_R$.
Then, the current through the device can be expressed as 
$I=(I_L+I_L)/2=(I_L-I_R)/2$.
The spin-flip process is neglected.
Although $\upa$-current and $\dna$-current have to be considered,
in this section, the suffix $s$ is omitted for the sake of convenience. 
The current formula is expressed by the element free Green's functions of QDs and electrodes. 
The Green's function of free QDs is given by
\begin{eqnarray}
g_{di}^<(\omega)&=& 2\pi i f_i(\omega) \delta (\omega-E_i),   \\
g_{di}^>(\omega)&=& 2\pi i [f_i(\omega)-1] \delta (\omega-E_i),  \\
g_{di}^r&=& \frac{1}{\omega-E_i+i\delta}, \\
g_{d2}^r(\omega)&=&\frac{1}{\omega-E_2+i\Gamma_{s}/2},\\
\Gamma_{s}&=&\Gamma_{Ls}+\Gamma_{Rs},
\end{eqnarray}
($i=1,3$).

The current $I_{Ls}$ provided in Ref.~\cite{TanaPRB} is expressed as 
\begin{eqnarray}
I_{Ls}&=&
\frac{e}{h}\int d\omega
\Bigl\{ 
\frac{\Gamma_{Rs} \Gamma_{Ls}}{|\Delta_c^r|^2D_2(\omega)}
 [f_{2L}(\omega)-f_{2R}(\omega)] 
\Bigr\},
\end{eqnarray}
where 
\begin{eqnarray}
D_{2s}(\omega)&=&(\omega-E_2)^2+(\Gamma_{Ls}+\Gamma_{Rs})^2/4,\\
C_{ij}&= &|W_{ij}|^2 g_{di}g_{dj}, \\
\Delta_c&=&
1-C_{12}-C_{32}.
\end{eqnarray}
Thus, the transmission probability is given by
\begin{eqnarray}
\mathcal{T}_{2s}(\omega) &=&
\frac{\Gamma_{Rs} \Gamma_{Ls}}{|\Delta_c^r|^2D_{2s}(\omega)}.
\end{eqnarray}

\subsection{Derivation of the Decoherence Time}
The decoherence time $t_{\rm dec}$ of the three QDs (Fig.1(b)) are the same as that 
provided in Ref.~\cite{TanaPRB}, and is expressed as
\begin{widetext}
\begin{eqnarray}
\lefteqn{
t_{\rm dec}^{-1}
\approx
\frac{ |W_{21}|^2 }{\hbar^2}
\Big\{
f(E_1) 
\left(
\frac{ F_2^>(E_1-\omega_{01})
}
{[E_1-\omega_{01}-E_2]^2+ \Gamma_s^2/4}  \right) 
+
[1-f(E_1)]
\left( 
\frac{F_2^<(E_1+\omega_{01})
}
{[E_1+\omega_{01}-E_2]^2+ \Gamma_s^2/4}   
\right)
\Big\}
}\nnm \\
&+& 
\frac{ |W_{32}|^2 }{\hbar^2}
\Big\{
f(E_3) 
\left(
\frac{F_2^>(E_2-\omega_{01})
}
{[E_2-\omega_{01}-E_2]^2+ \Gamma_s^2/4}  \right)
+
[1-f(E_3)]
\left(
\frac{F_2^<(E_2+\omega_{01}) 
}
{[E_2+\omega_{01}-E_2]^2+ \Gamma_s^2/4}  \right) 
\Big\},
\label{tdec3}
\end{eqnarray}
where $\omega_{01}=|E_1-E_2|$ and,
\begin{eqnarray}
F_2^<(\omega)&=&\Gamma_{Ls} f_L (\omega) +\Gamma_{Rs} f_R(\omega),\\
F_2^>(\omega)&=&\Gamma_{Ls} [1-f_L (\omega)] +\Gamma_{Rs} [1-f_R(\omega)].
\end{eqnarray}
Here, $f(E_i)$ and $1-f(E_i)$ indicate the existence of an electron at the $E_i$ level 
and no electron at the $E_i$ level ($i=1,3$), respectively.
For the average case, $f(E_i)=1/2$ is considered in the following calculations.

\end{widetext}

\section{Temperature Dependence and the Case of $L=10~\mu$m}\label{Temp}
\begin{figure}
\centering
\includegraphics[width=8.0cm]{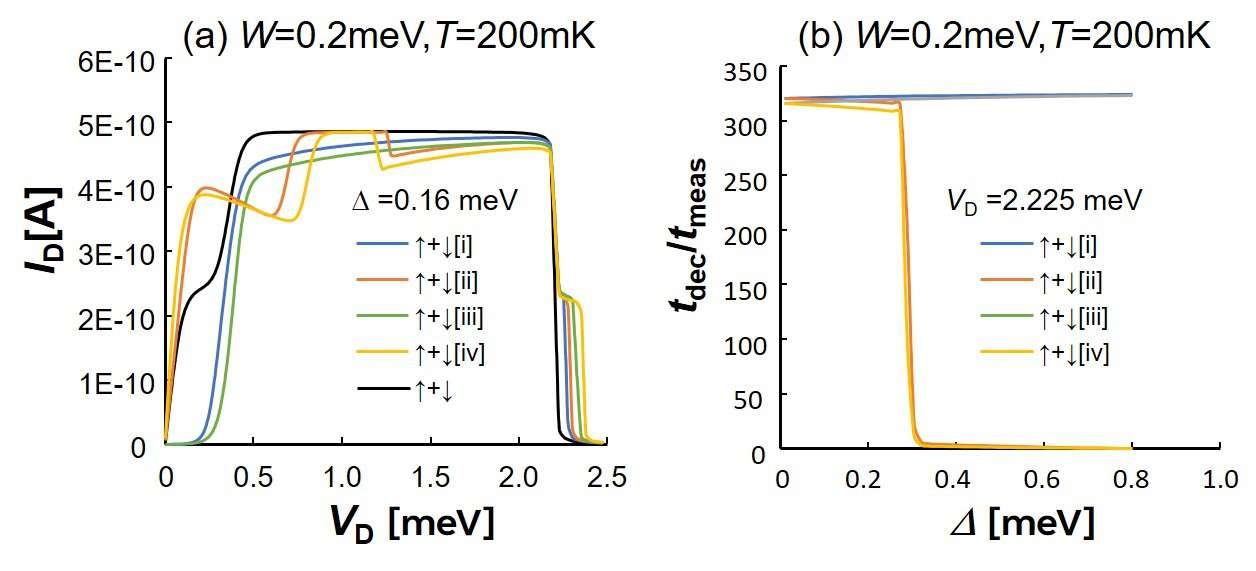}
\caption{
(a) $I_D$-$V_D$ characteristics and (b) $t_{\rm dec}/t_{\rm meas}$ 
at $T=200$ mK, $W=0.2$ meV, and $V_{\rm g}-V_{\rm th}=0.1$ V. 
Other parameters are the same as those considered for Fig.~\ref{figSQD2}.
}
\label{figt}
\end{figure}
Figure~\ref{figt} illustrates the number of 
possible measurements at $T=200$ mK, 
which is compared to the results shown in Fig.~\ref{figSQD2} in the main text.
As the temperature increases, the resonant peak blurs and the sensitivity weakens. 
Thus, the operating temperature is important while performing the measurements.

\begin{figure}
\centering
\includegraphics[width=8.0cm]{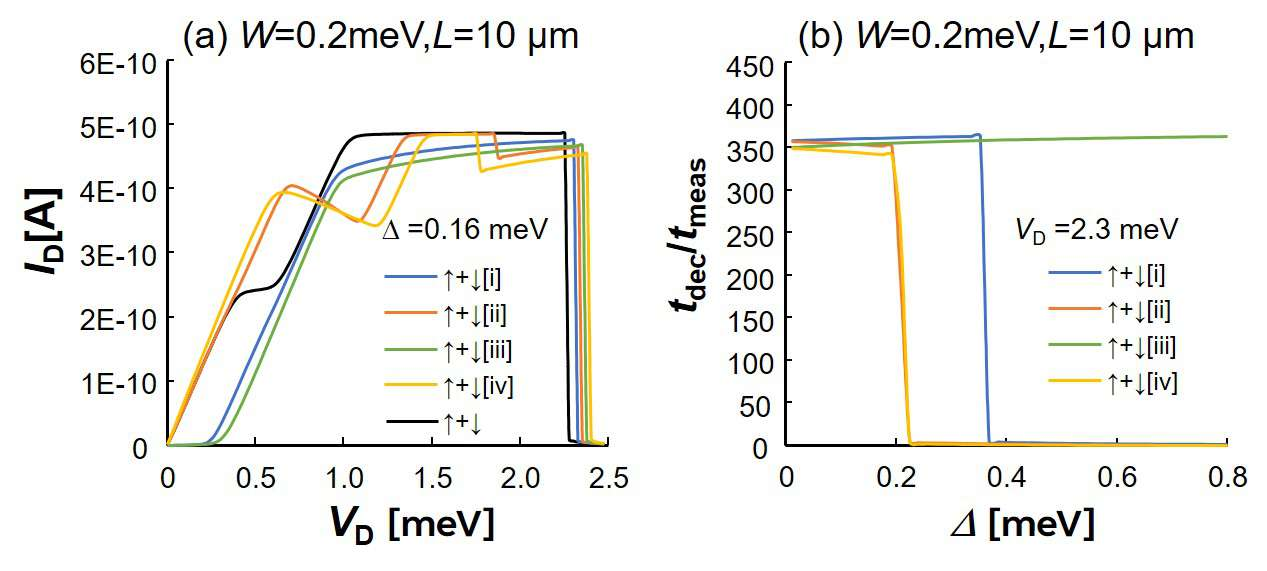}
\caption{
(a) $I_D$-$V_D$ characteristics and (b) $t_{\rm dec}/t_{\rm meas}$ 
of $L=10~\mu$m at $T=100$ mK. 
$\Gamma_0=2\mu$eV, and $V_{\rm g}-V_{\rm th}=0.1$ eV.
Other parameters are the same as those considered for Fig.~\ref{figSQD2}.
}
\label{figl1000nm}
\end{figure}

Figure~\ref{figl1000nm} shows the $I_D$–$V_D$ characteristics and 
$t_{\rm meas}/t_{\rm dec}$, 
where the gate length of the transistor is given by $L=10\mu$m with $W=0.2$~meV 
and $V_{\rm g}-V_{\rm th}=0.1$~eV.
On comparing Fig.~\ref{fig4} and Fig.~\ref{figSQD2}, 
the $I_D$–$V_D$ characteristics are observed to have changed 
owing to the large resistance of the transistor. 
 
\section{Two-dimensional Qubit Array with Common Gate Structure}\label{2D}
\begin{figure}
\centering
\includegraphics[width=8.5cm]{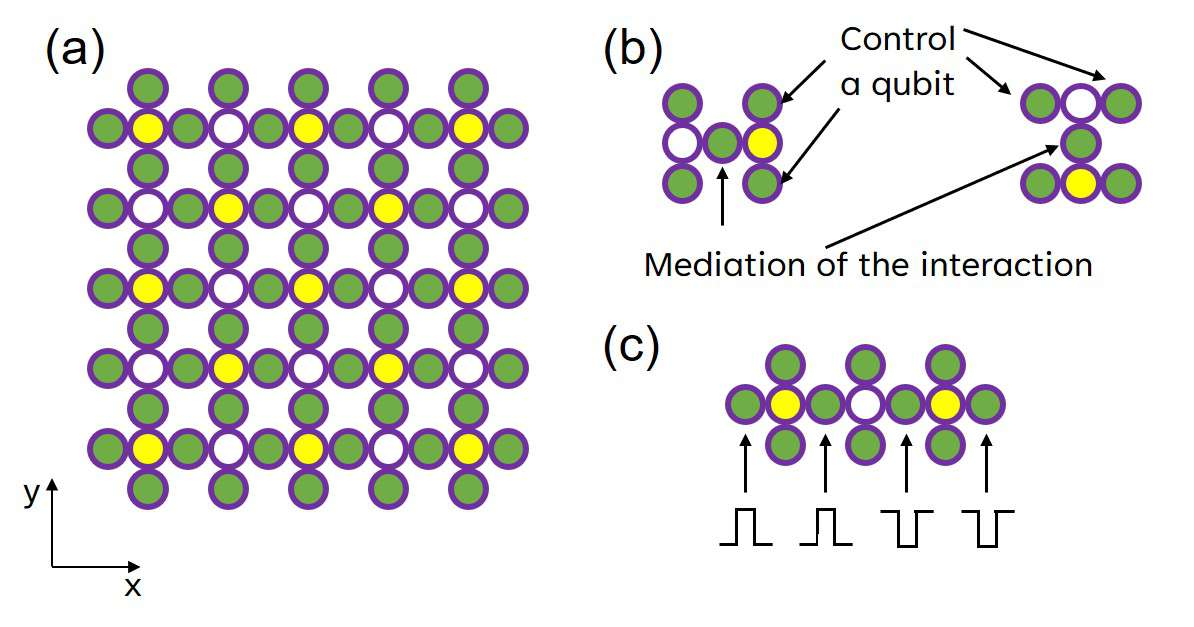}
\caption{
(a) A two-dimensional qubit array as an extension of Fig.1 in the main text.
(b) Qubit-qubit operations are conducted in both x and y-directions. 
The channel-QDs work as the controlling line to mediated interactions 
between the qubits and for the measurement of two adjacent qubits. 
(c) Different phase pulses used to avoid undesirable operations.
}
\label{figArray}
\end{figure}

A two-dimensional qubit array consisting of the units shown in Fig.1 of the main text is considered.
Figure~\ref{figArray}(a) shows the schematic of a qubit array.
A single qubit operation is assumed to have been performed 
using the magnetic-field gradient produced by the micromagnets around the array~\cite{Noiri1,Noiri}.
The channel-QDs (green circles) are shared by two qubits (yellow and white circles), 
such that the changes in the electric potential of the channel-QDs affect both sides of the qubits (crosstalk), 
as shown in Fig.~\ref{figArray}(b).
In other word, each spin direction is changed by the two surrounding nearest channel-QDs. 
For example, by changing the pulse voltage on the nearest channel-QDs, 
the qubit sandwiched between the two channel-QDs experiences an oscillating field, 
which results in a spin direction change under the magnetic-field gradient. 
Because each channel-QD is shared with two qubits, 
when a pulse voltage is applied to the channel-QD, 
the voltage change affects the qubits on both sides of the channel-QDs.
Thus, the random-access process is not suitable in this structure.
However, the cross-talk can be avoided, as shown in Fig.~\ref{figArray}(c).
By applying different phase pulses to the nearest channel-QDs, 
the effect of the applied pulse 
between the center qubits can be negated.

When the channel-QDs are used as the control line to change the spin direction, 
as shown in Fig.~\ref{figArray}(a) and (b),
the resistance between the QD and the source and drain is greater than 50 $\Omega$ 
owing to the tunneling resistance between the channel-QD and electrodes. 
Thus, a high-frequency signal does not pass through  
the channel-QD and is reflected.
It produces an additional electric field to the qubits, 
as if the channel-QD component functions as the gate electrode. 
Detailed discussion on the operations are beyond the scope of this study.

\begin{figure}
\centering
\includegraphics[width=8.5cm]{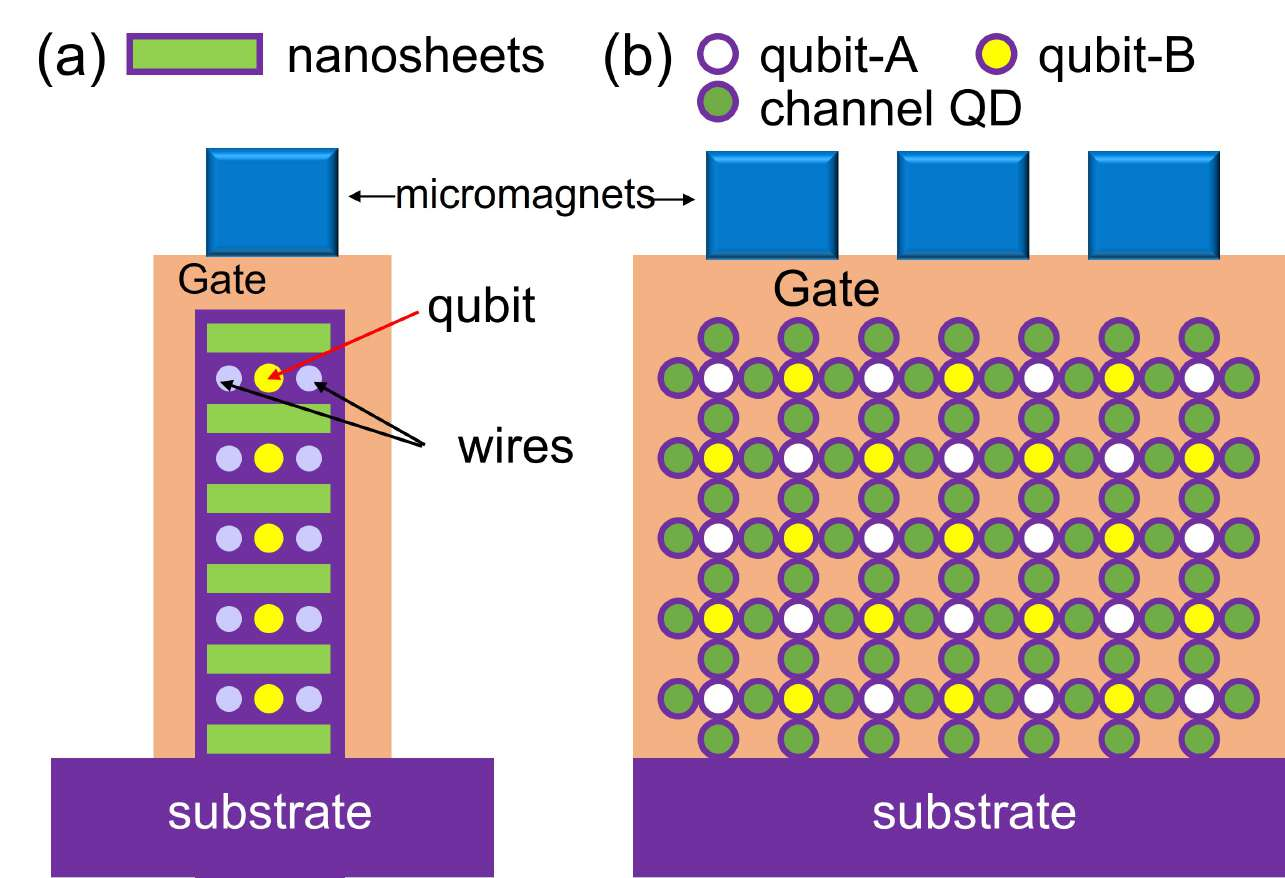}
\caption{
Cross-section of the stacked qubits.
(a) Qubit system embedded into nanosheet as a modification of the system discussed in Ref.~\cite{TanaAIP}.
(b) As an extension of (a), toy system of two-dimensional qubit array.
The system shown in Figure~\ref{figArray} is stacked on the substrate 
using a common gate, which controls the potential of $E_2$, shown in Fig.1 in the main text.
The qubit-QDs and the channel-QDs are placed side by side.
The qubits A and B have the same structure.
}
\label{fig3D1}
\end{figure}
\begin{figure}
\centering
\includegraphics[width=8.0cm]{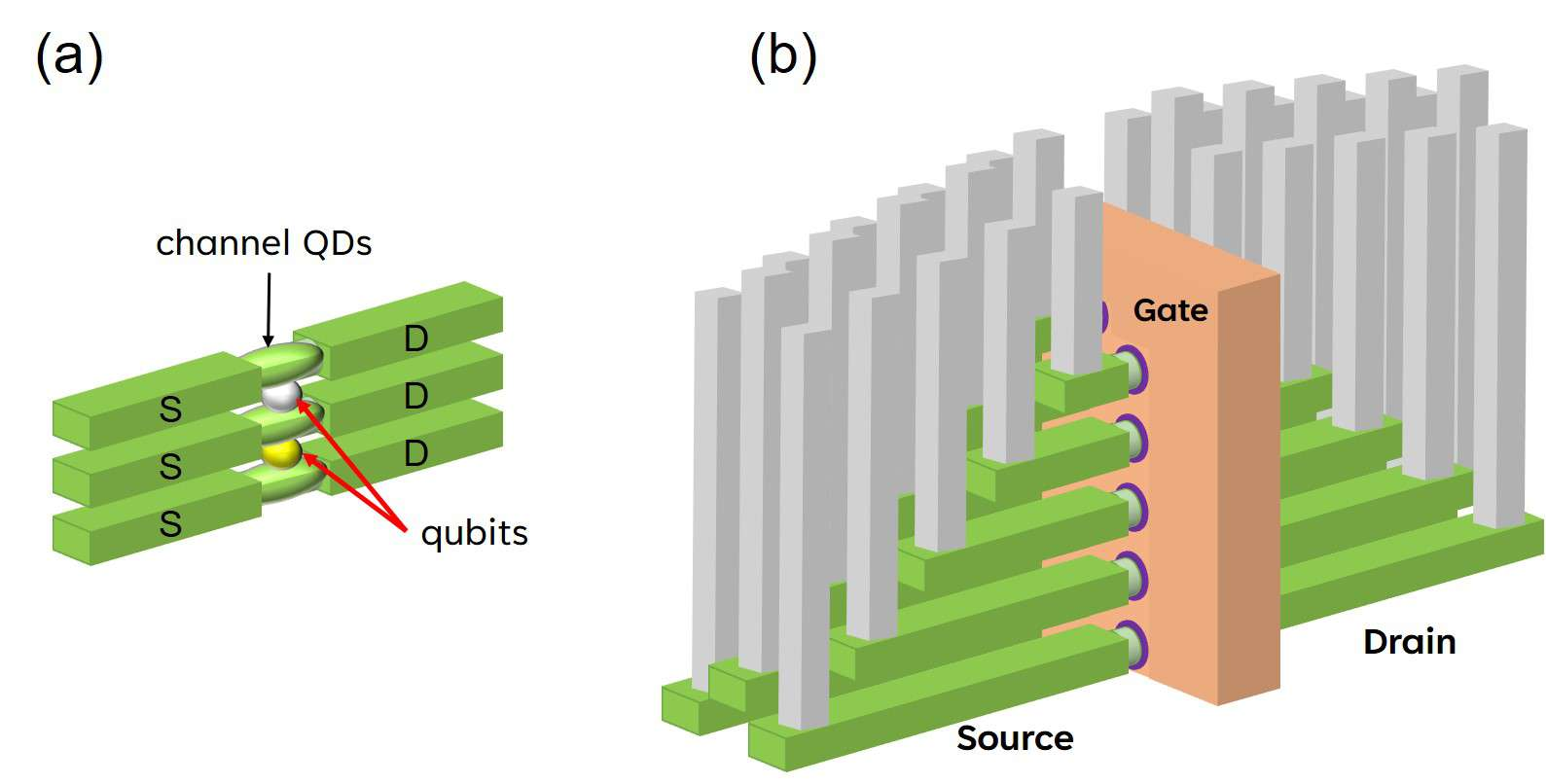}
\caption{
Bird's eye view of the 3D stacked qubit system.
(a) Qubits are surrounded by channel-QDs. 
(b) Sources and drains of different channel-QDs are independently accessed by 
attaching different bias lines to the sources and drains.
}
\label{fig3D2}
\end{figure}

\section{Common Gate Stacked Qubit Structure}\label{3D}
Here, we would like to propose a stacked qubit structure as a typical embodiment of Fig.~\ref{figArray}.
Particularly, the common gate concept is applied to the nanosheet 
to reduce the number of wires required in the system. 
The nanosheet is one of the most advanced semiconductor structures available, 
where the channel parts are stacked on the substrate.
Qubits are represented by QDs embedded between the channels, as described in Ref.~\cite{TanaAIP}.
Figure~\ref{fig3D1} shows a cross-section of the proposed stacked qubits as an extension of the nanosheet. 
The gate electrode shared by the channel-QDs controls the electric potential of the QD system.
Figure~\ref{fig3D2}(a) shows a bird's eye view of the stacked qubit system.

Figure~\ref{fig3D1}(a) shows a simple extension of the system discussed in Ref.~\cite{TanaAIP}, 
which is rotated by 90 degree of the previous proposal using a FinFET structure.
The qubits are embedded into the oxide layer between the two channels.
Here, the magnetic gradient method is applied and the two nearest metallic wires are used 
to generate an oscillating electric field gradient on the electrons in the qubit.
In this structure, the distance between qubits in the lateral direction is still larger than that in the 
vertical direction; therefore, the lateral coupling is weaker than the vertical coupling.
Figure~\ref{fig3D1}(b) shows the advanced structure  
in which the coupling magnitude of the lateral qubits is equal to that of the vertical coupling. 
The channel-QDs play both the roles of detection 
and interaction between the nearest qubits, as discussed in Fig.~\ref{figArray}.

Figure~\ref{fig3D2} (a) shows the bird's eye view of the 3D stacked qubit system.
The source and drain of each channel-QDs are separated to independently control the Fermi energies
of the QDs. 
The interaction of the qubits is switched on and off according to the position of the Fermi energy.
The size of the contact to each source and drain should be sufficiently small.
Such a structure is different from that of a typical nanosheet but 
its feasibility will be high when fabrication technologies for 3D NAND flash memories are applied. 
Although the case in which the channel consists of a QD has been discussed,
nanowires can replace the center-QD.

\begin{figure}
\centering
\includegraphics[width=5.0cm]{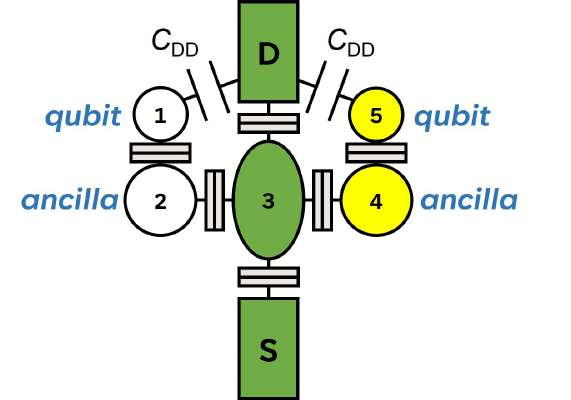}
\caption{
Extended version of the qubit system in which 
additional ancilla qubits (QD2 and QD4) are placed between the 
channel-QD(QD3) and target qubits (QD1 and QD5) to reduce the effect of backaction.
If the target qubit size is smaller than the ancilla qubits,
the tunneling of electrons between the target qubits and the electrodes 
is prohibited. In addition, because the target qubits are set close to the electrodes, 
their energy levels can be controlled by the potential of the electrode
relative to the common gate.
Due to the different sizes of the QDs, the energy levels are different.
}
\label{fig2operation}
\end{figure}

\section{Five QD Operation}\label{FiveQD}
Yoneda {\it et al. }~\cite{Yoneda} proposed that ancilla QD enhances the coherence of target qubits. 
Hence, we can also consider the ancilla QD between 
the channel-QD and target qubits.
Figure~\ref{fig2operation} shows the schematic of the 5-QD case, 
where the ancilla QDs (QD2 and QD4) are placed 
between the center-QD (QD3) and the target qubits (QD1 and QD5).
The target qubits, which are placed close to the electrodes, are smaller than the ancilla QDs.
This leads to stronger capacitive coupling between the target
qubits and electrodes where no tunneling between the qubits and electrodes is assumed. 
As QD1 and QD5 are close to the electrodes, 
their potentials can change significantly by the applied bias, 
as compared with those of the ancilla QD2 and QD4.  
The electrons are introduced into QD1 and QD5 by lowering 
the potentials of the electrodes to a value lower than that of the 
channel-QD controlled by the common gate.
When a large bias is applied, to avoid current flow through the channel-QD, 
the S/D voltages are set to be equal.
Detailed discussion on the operations are beyond the scope of this study.


\end{document}